\documentclass[conference]{IEEEtran}
\IEEEoverridecommandlockouts
\usepackage{cite}
\usepackage{amsmath,amssymb,amsfonts}
\usepackage{algorithmic}
\usepackage{graphicx}
\usepackage{textcomp}
\usepackage{xcolor}
\usepackage{subcaption}
\usepackage{graphicx}
\usepackage{comment}
\def\BibTeX{{\rm B\kern-.05em{\sc i\kern-.025em b}\kern-.08em
    T\kern-.1667em\lower.7ex\hbox{E}\kern-.125emX}}

\begin{document}

\title{ 
Decentralized P2P Trading based on Blockchain for Retail Electricity Markets
}

\author{
  \IEEEauthorblockN{Masoud H. Nazari and Antar Kumar Biswas}
  \IEEEauthorblockA{Department of Electrical and Computer Engineering, Wayne State University, Detroit, Michigan \\
    masoud.nazari@wayne.edu and hr2122@wayne.edu}
}

\maketitle

\begin{abstract}
This paper introduces peer to peer (P2P) trading mechanisms based on decentralized Blockchain to facilitate retail electricity market for ever-increasing distributed energy resources (DERs). The Blockchain network supports fast and secure retail trading among DERs and facilitates a sustainable local P2P trading platform. In this decentralized Blockchain architecture no single entity or organization has control over the entire system rather all users collectively maintain control. 
The effectiveness of the proposed automated market design and optimization is simulated using different use case scenarios in an open source Blockchain Simulator and MATLAB. The results show the efficacy of the trading mechanism in achieving demand response through strategies such as peak load shaving, load shifting, and integration of DERs.


\end{abstract}


\section{Introduction}\label{sec:intro}


 Nowadays, researchers from various disciplines, such as real estate
~\cite{mizrahi2015blockchain}, healthcare~\cite{kar2016estonian}, utilities~\cite{lacey2016energy}, finance~\cite{kelly2016forty}, and the government sector~\cite{walport2016distributed}, are interested in the Blockchain technology. 
The Blockchain mechanism has a secure computerized 
framework that supports millions of energy trading transactions, such as prosumer (seller) to 
consumer (buyer) or prosumer to utility. The users can trade energy with each other 
without the intervention of any traditional energy distributors. 
Blockchain can formulate a decentralization of P2P trading while adding security. This ensures the integrity of the data transaction traceability.
The Blockchain mechanism with a decentralized structure offers numerous benefits, such fast transactions and privacy. The Blockchain P2P energy trading network is modeled and implemented using Hyperledger \footnote{Hyperledger is an open-source collaborative effort aimed at advancing cross-industry Blockchain technologies. It is not a single Blockchain network but rather a collection of open-source projects and tools.}~\cite{cachin2016architecture}. 

This paper first offers a comprehensive background on the P2P trading platform and P2P energy trading architectures. P2P energy trading is  referred to trading between prosumers and consumers. The peers can trade energy with each other without the intervention of any traditional energy distributors.
Over the past few years, P2P trading has gained significance as an emerging future energy market strategy for smart community grids.
In this trading mechanism, prosumers can prosper by selling their excess energy or by participating in demand response (DR) programs 
(for instance, reducing HVAC loads, household load scheduling, etc.). Similarly, utilities and network providers gain satisfaction in terms of high system reliability with reduced peak demand~\cite{khalid2020blockchain}, minimized reserve requirements, and lower operational and investment costs. In a nutshell,  P2P trading is beneficial for prosumers, utilities, and network providers. 
However, the integration of DERs into the present local electricity markets and incorporation of P2P trading can be challenging due to the complex and often centralized nature of traditional energy systems. 

In ~\cite{zhou2021credit}, authors investigate how the Blockchain technology can be used to support P2P energy trading within DERs. A DR platform based on game theory has been developed in \cite{9542944} to facilitate efficient and secure energy trading among prosumers.
This paper presents, a new decentralized Blockchain-based P2P trading mechanism for retail electricity markets. Main contributions of the paper are as follows:

1) We use a Blockchain-based P2P market design and
optimization with a focus to determine optimal entry and exit energy trading windows using user inputs, environmental information, and supervised learning.

2) The proposed Blockchain-based P2P trading network is modeled using Hyperledger. With the Blockchain Hyperledger platform, 3000 to 20,000 transactions can be performed per second\cite{dreyer2020performance}, which makes it an effective candidate for real-time retail electricity markets.

3) We demonstrate how the Blockchain-based P2P trading helps providing DR services 
and reducing overall energy consumption.


The remaining of the paper are summarized as follows: 
In Sections II we introduce a comprehensive background of P2P Trading along with key P2P techniques. In section III we introduce the Blockchain technology and Blockchain trading strategies. In Section IV, we discuss the experimental results and simulation studies. The overall conclusion and future scope are summarized in Section V.

\section{Peer to Peer Energy Trading }\label{sec:background}

\subsection{Key Elements}
\begin{figure}[!ht]
    \centering
    \includegraphics[width=.3\textwidth]{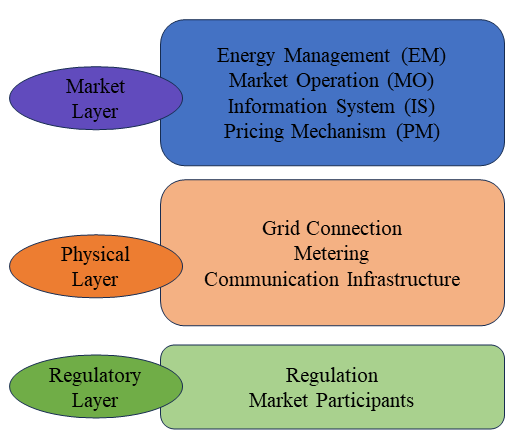}
    \caption{P2P network layers.}
    \label{fig:P2P key layers}
\end{figure}

In a P2P network, trading can take place without an intermediary~\cite{sousa2019peer}. This offers flexibility to add or remove peers at any instant in time. Particularly, in case of network service interruptions or losses in the system \cite{MN1, MN2}. The entire network is divided into three layers, namely the market layer, physical layer, and regulatory layer~\cite{tushar2018transforming}, as shown in Fig.~\ref{fig:P2P key layers}.

\subsubsection{Market Layer}
The market layer provides a secure platform for all users to share information and resources within the network~\cite{tushar2018transforming}. All financial settlements take place over this layer, wherein bids and offers are created, and matched based on market-appropriate mechanisms~\cite{tushar2020peer}. 

    
    
    

\subsubsection{Physical Layer}

Once the financial transactions are completed over the market layer, the actual power transfer takes place over the physical layer~\cite{tushar2020peer}. 

    
    


\subsubsection{Regulatory Layer}
For P2P trading to be possible, there should be an adequate number of market users and supporting regulatory policies, which are the elements of this layer~\cite{tushar2020peer}.

    

\subsection{Decentralized Market Architecture
}


\begin{figure}[!ht]
    \centering
    \includegraphics[width=.45\textwidth]{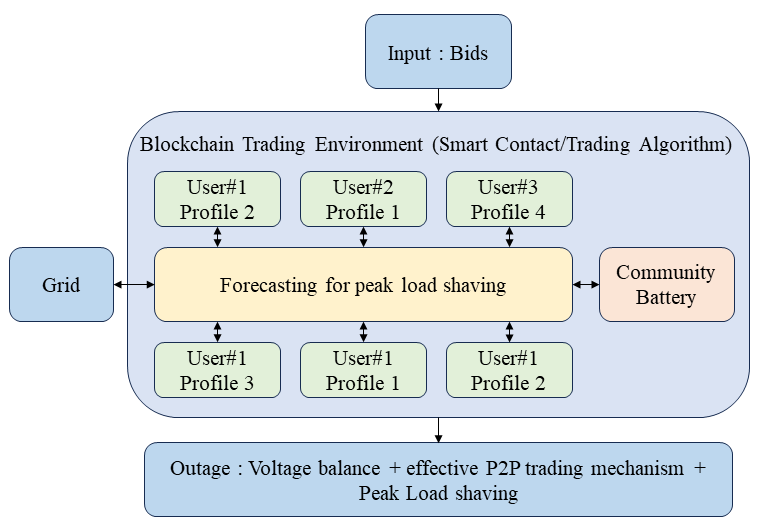}
    \caption{ Decentralized P2P energy trading architecture.}
    \label{fig:market_architecture}
    \end{figure}

    
    
    A decentralized market enables direct communication among peers. Both trading and data communication are done in a decentralized way~\cite{morstyn2018bilateral}. This structure offers freedom and flexibility for peers to decide whether to take part in energy trading at any given instant of time while maintaining privacy. Exceptional scalability is also an advantage~\cite{tushar2020peer}. On the other hand, the absence of a central party limits the security of the market which can impact user participation.

\subsection{P2P Trading Methods}

It is evident that DER integration and P2P trading implementation can be challenging. This paper focuses on the market layer's challenges, intending to determine possible solutions to resolve these issues on the market layer.  

The key techniques for P2P implementation include: 1) Cooperative and non-cooperative games to derive an optimal solution on mutual benefit of both seller and buyer, 2) Auction theory \cite{thomas2019general}, which derives optimal solution from the point of intersection, 3) Constrained optimization
\cite{morstyn2018multiclass}
that implements the mathematically programmed technique to optimize P2P trade parameters under a variety of market constraints. Blockchain 
\cite{tushar2020peer}
adds an extra secure platform, where each new user and every new transaction needs to be verified. This can ultimately attracts more prosumers and consumers and thereby reduce overall costs. The key highlight of the Blockchain mechanism is the decentralized structure, where trading and data transfer can operate without a central party but achieve the same functionality with the same amount of certainty \cite{faizan2019decentralized}. 
Therefore, the Blockchain mechanism can be an effective candidate for real-time { market design and
optimization}.

\section{Blockchain Technology}

Blockchain is a technology that stores data in a block. After every transaction a new block is generated which is linked with the previous block. This creates a chain of blocks that operates in a decentralized fashion. Blockchain is based on Distributed Ledger Technology (DLT). DLT relies on a consensus and 
communication protocol that safeguards the ledger’s integrity through connected 
cryptographically time-stamp block that represents transactions \cite{mizrahi2015blockchain}. In Blockchain, each block contains sender/prosumer and receiver/consumer information, details of the amount of energy that needs to be transferred, and most importantly a unique identification number called ``hash" with time stamp 
\cite{peck2017energy}. Manipulation of data on any block can invalidate the hash of all the following blocks, which makes it very robust and secure against hacking. Moreover, Blockchain does not expose a single point of failure and thereby adds security and improves privacy 
\cite{wang2019energy},\cite{li2019design}. 

\begin{figure}[!ht]
    \centering
    \includegraphics[width=.45\textwidth]{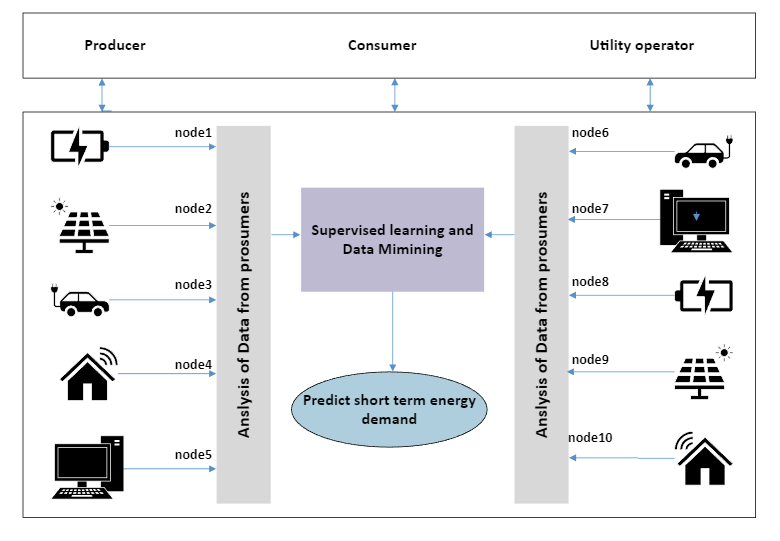}
    \caption{Blockchain energy trading architecture.}
    \label{fig: predictive netwrok}
\end{figure}

\begin{figure*}[!ht]
    \centering
    \includegraphics[width=1\textwidth]{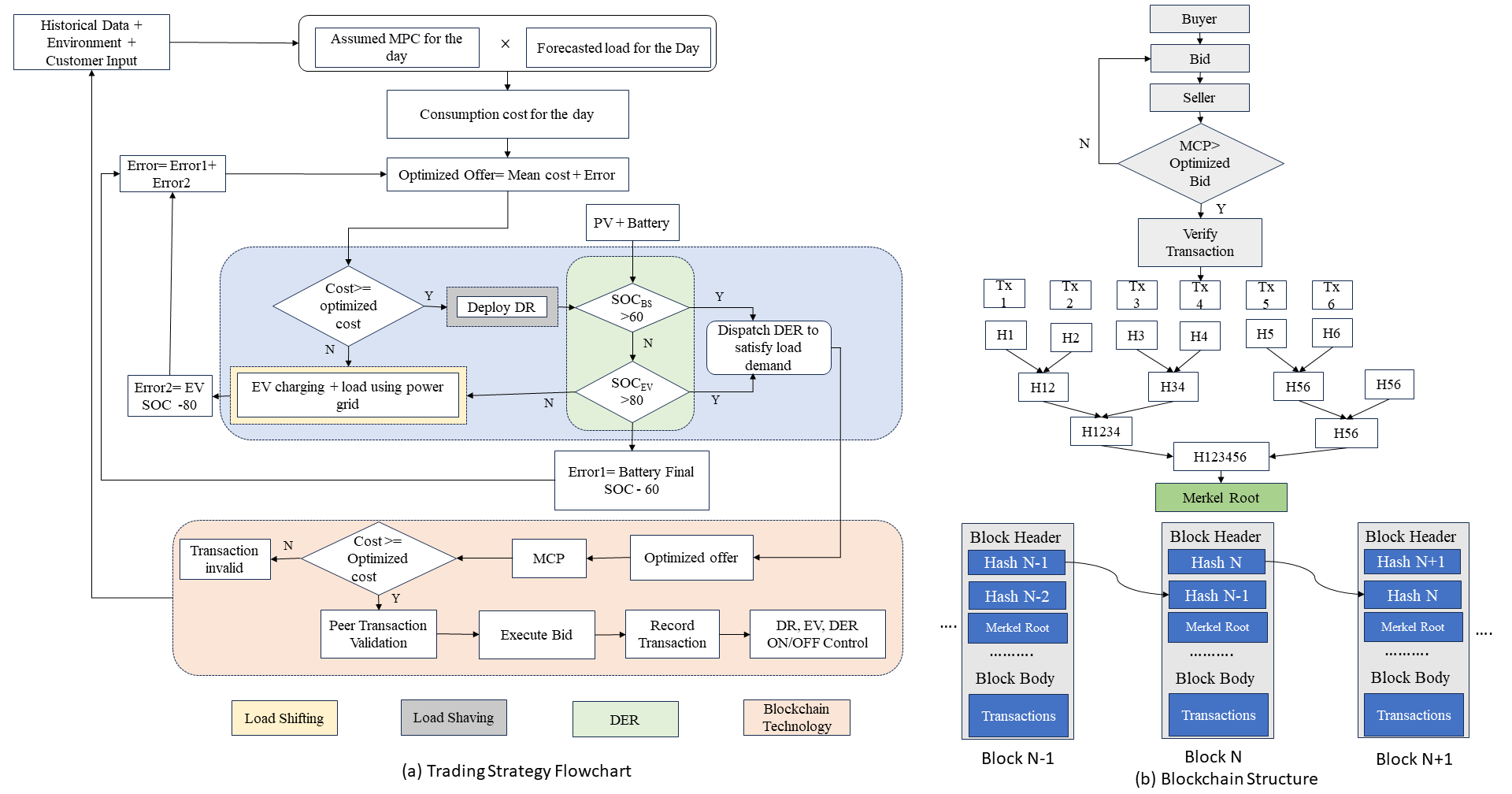}
    \caption{{ Blockchain market design and optimization algorithm.}}
    \label{fig:trading_strategy_flowchart}
\end{figure*}


There are different types of Blockchain network including: Hyperledger, Ethereum, Tezos, EOSIO, Stellar, Quorum, etc. Depending on the platform, the transaction time can vary. Considering P2P trading properties, it was identified that Blockchain Hyperledger would be a suitable option \cite{gorenflo2020fastfabric}. 
The Linux Foundation hosts the Hyperledger project.

\subsection{Blockchain Trading Architecture}
 The {Blockchain market design and optimization} architecture includes two modules:
  1) The predictive Blockchain-based P2P energy trading, and
  2) The intelligent energy prediction model to fulfill energy
  demand in a distributed network. The first module focuses on using Blockchain for P2P trading, while the second module is concerned with creating a forecasting model for effectively meeting demand.
  

  The energy consumption data from DERs supported by this architecture are analyzed using different supervised learning and data mining approaches to meet future energy demand. The pre-processed data is used for real-time and day-ahead scheduling of DERs. Moreover, the time-series features, such as hourly, daily, and weekly are used to predict the short-term energy demand using supervised learning models. 
    The participants, such as prosumer, consumer, and
  utility operator, can interact through the Blockchain architecture (Fig.~\ref{fig: predictive netwrok}, which is used for secure energy trading.

{
\section{Simulation Results} }\label{sec:sec5}
The focus of this paper is to develop a prosumer trading strategy by reducing energy consumption, maximizing cost savings, and making local P2P trading network reality. Since the available energy at the peer node is limited to stored energy and power generated by the photovoltaic (PV) array, it is essential to know { when one participates in energy trading or continues to use energy from the grid.} Hence, there is a need for an optimized prosumer strategy to plan the entry and exit of a trade. 
The optimized strategy could be deployed as a phone application wherein the customer can share day ahead inputs and/or use environmental data for calculating optimized bid/offer to prepare the P2P trading system. For example, if the prosumer is planning to do a task that requires significant electrical energy, its energy management system can queue in this task to the phone application. The trading system would then use other environmental information like weather data to estimate heating/cooling needs or to estimate available solar energy, project loads/market-clearing cost, and recommend the prosumer an optimized slot for completing the task. Although it is a difficult task to estimate load consumption on a real-time basis, by scaling this load usage strategy across a small size local P2P trading forecasting loads and managing P2P trading within a local energy market would be beneficial. Furthermore, with the help of supervised learning, the forecasted market clearing price (MCP), load trend, and thus the prosumer load consumption cost vs time of the day can be estimated {\cite{MN2023}}. 

The detailed steps of the Blockchain trading algorithm (Fig.~\ref{fig:trading_strategy_flowchart}(a)) are as follows:   

\begin{itemize}
    \item Step 1: Derive estimated consumption and MCP for the day using user inputs and supervised learning. The user could be a prosumer or a consumer. 
    
    \item Step 2: Calculate optimized bid/offer for the particular market interval as needed. For the first pass of the optimization, the strategy uses the average estimated consumption cost from Step 1.

    \item Step 3:  Check if MCP is greater than the optimized bid/offer. {If MCP is greater, the strategy would be first responding to the demand by lowering energy consumption to a level set by the prosumer.} If not, the strategy would be to use the power grid to support loads. 
    
    \item Step 4: Check for battery as in (1) and electric vehicle (EV) state of charge (SOC) as in (3), conditions for optimized energy usage. 
    
\begin{align}
    &SoC^{min}_{BS} \leq SoC_{BS}(t) \leq SoC^{max}_{BS}  \\
    &SoC^{min}_{BS @ t\_end} = 60  \\
    &SoC^{min}_{EV} \leq SoC_{EV}(t) \leq SoC^{max}_{EV} \\
    &SoC^{min}_{EV @ t\_end} = 80 
\end{align}

{ where, $SoC^{min}_{BS}$, $SoC^{max}_{BS}$, $SoC^{min}_{EV}$ and $SoC^{max}_{EV}$ represent minimum and maximum SOC for battery and EV respectively. To be eligible for trading, the minimum SOC of a battery has to be 60\%, and an EV's minimum SOC must be 80\% which are represented by (2) and (4).}
    \item Step 5: Estimate battery and EV SOC error and obtain optimized bid/offer before the start of the market trading interval using Newton Raphson method. 

    \item Step 6: Enter Blockchain trading wherein the offer is validated at participating peer nodes, executed, recorded, and appropriate ON/OFF switch actions are taken at respective DR/EV/DER locations.

    \end{itemize}

{ In Fig.~\ref{fig:trading_strategy_flowchart}(b), the creation of Block and Blockchain is presented. When the Minimum Compatible Price (MCP) exceeds the optimized bid and a transaction is verified by endorsing nodes within the network, the transaction is confirmed and a unique hash is created to represent the transaction. These individual transaction hashes are paired and undergo additional hashing, continuing until a singular root hash is generated, known as the Merkle root. Subsequently, this Merkle root is integrated into the Block header.
In addition to the Merkle root, the Block header contains the hash of the current Block, alongside the hash of the preceding block, establishing an interconnected sequence known as the Blockchain.} 




 The intelligent energy prediction model uses Blockhain which consists of real-time and day-ahead energy trading based on pre-processed data and short-term energy prediction.

\begin{table}[!ht]
\begin{center}
\begin{tabular}{|c|c|}
\hline
\textbf{Simulation Parameters} & \textbf{Input} \\ \hline\hline
Number of Solar panels & 1 \\ \hline
Solar data input & 2D look table \\ \hline
Load data input & 2D look table \\ \hline
Simulation time & 3600 seconds \\ \hline
Home Battery Capacity & 250 KWh \\ \hline
Initial Home Battery SOC & 60 \% \\ \hline
EV Battery Capacity & 80 KWh \\ \hline
Initial EV SOC & 50 \% \\ \hline
Final EV SOC at 6 am & 80 \% \\ \hline
Demand Response & \begin{tabular}[c]{@{}c@{}}10\% reduction during peak\\ projected electricity cost\\ consumption\end{tabular} \\ \hline
\end{tabular}
\end{center}
\end{table}

 In a conceptual scenario, each node of the proposed intelligent energy platform is utilized to store and process energy trading data. For simplicity, the simulation is limited to a single prosumer that would have access to two energy sources, grid and distributed solar generation using energy storage system, and DR program supported by the local utility. Simulation is done using { an open source Blockchain simulator \cite{BlockDem}} and MATLAB. Table 1 indicates the input data and simulation parameters. Fig. 5 lists some of the key events that include EV load shifting, power distribution throughout the day, and loads supported by the battery during peak demand.




Subplot 5-a represents the estimated MCP that is an input to Step 1. { The MCP is calculated using data stored in Hyperledger Blocks.} Subplot 5-b shows power distribution. The key items to consider are solar power rise at dawn, peaks at maximum irradiates power available, and settles at dusk. { In instances where the MCP falls below the optimized bid, it becomes more advantageous to leverage the power grid to meet the demand. To optimize energy consumption, EV load shifting strategies are implemented, allowing for charging activities to occur during periods characterized by the lowest electricity consumption cost.}
 
Positive power indicates flow from the grid to load and vice versa. {For instance, it can be observed that the power is negative during the peak load.} Items 2, 3, and 4 are part of the strategy to achieve peak load shavings on the grid side and to maximize prosumer earnings.

Similarly, subplot 5-c represents home battery storage and EV SOC indicator. Note that the EV SOC level is required to be at least 80\%  by 6 AM to {meet the user needs}. Also, the EV consumption is indicated by SOC dropping down to 50\% at 4 PM. Subplot 5-d shows the overlay of load consumption with actual time instants where the load is supported by DER and the DR program. 
The plot also shows the cost of energy consumption using the grid, user-defined offer for solar power utilization 
and actual user consumption cost. 


    \begin{figure*}[!t]
     \centering
     \begin{minipage}[t]{0.49\textwidth}
         \centering
         \includegraphics[width=\textwidth]{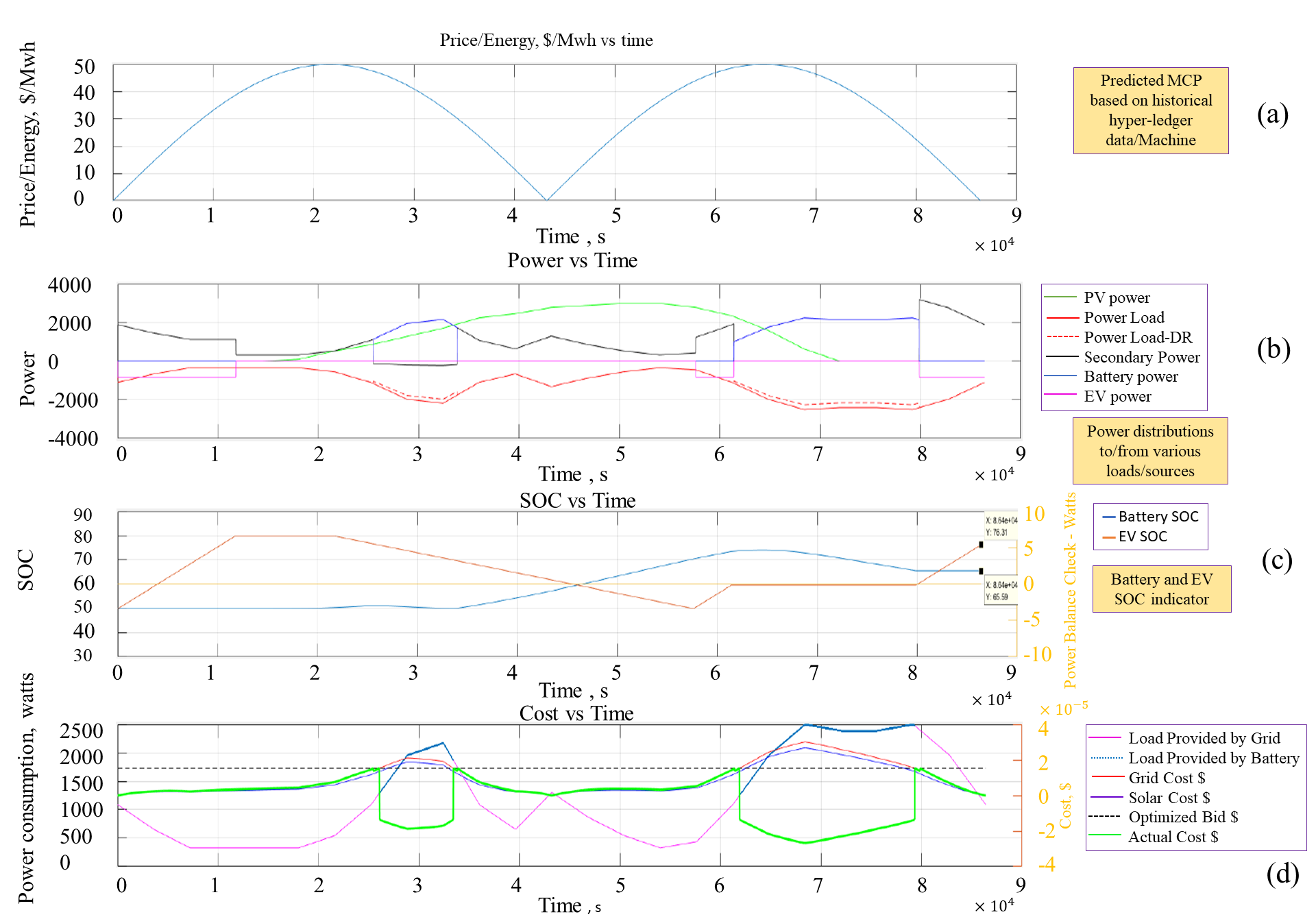}
         \caption{{Simulation results, initial optimization iteration.}}
         \label{first}
     \end{minipage}
     \hfill
     \begin{minipage}[t]{0.49\textwidth}
     \centering
     \includegraphics[width=1\textwidth]{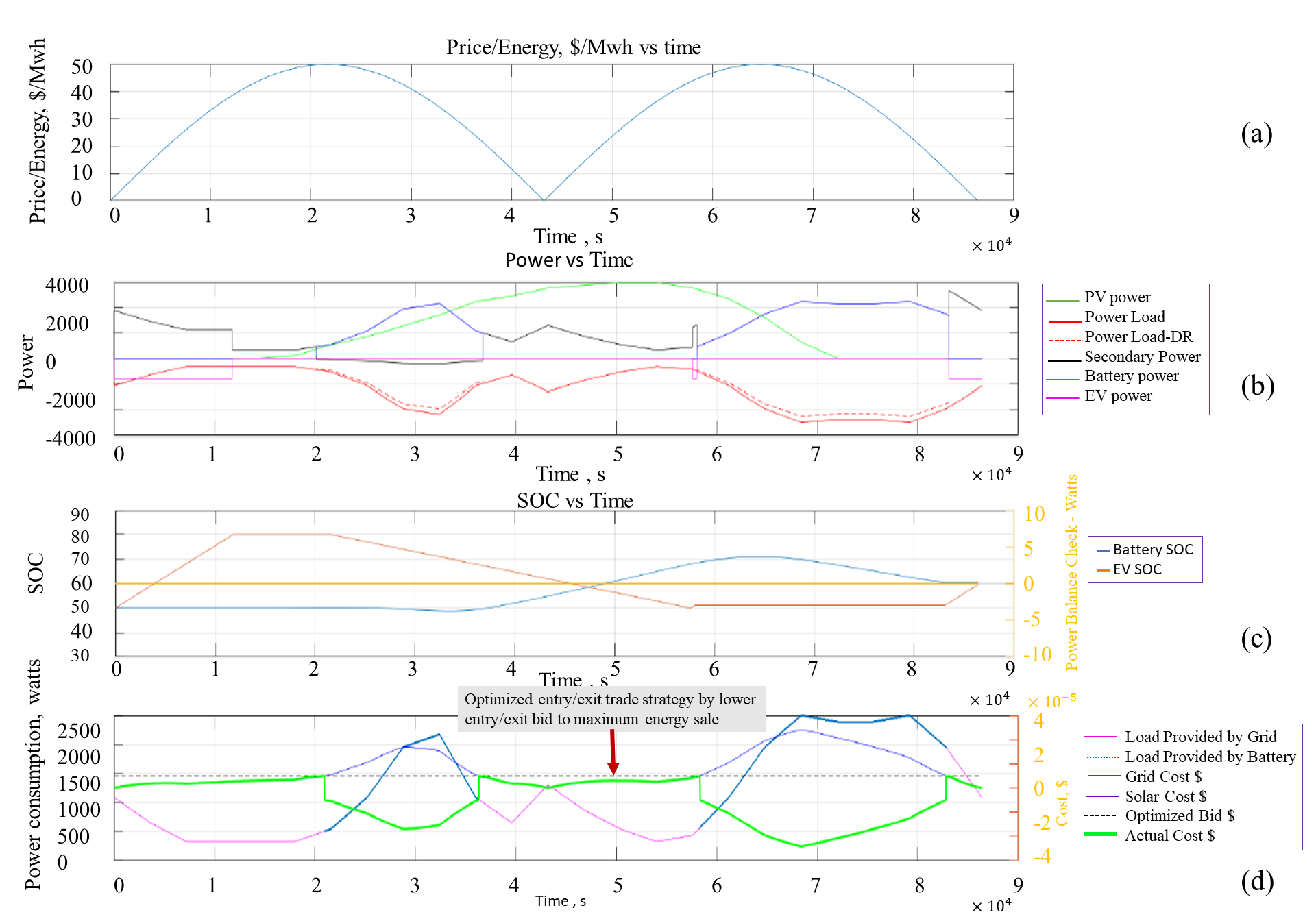}
     \caption{{Simulation results, final optimization iteration.}}
    \label{final}
     \end{minipage}
     \end{figure*}


Fig. \ref{final} shows the results of the final iteration. The highlighted window indicates trade entry and exit points. {The timing for this window is critical to maximize earnings and prepare DER to support the transaction.} For example, {the strategy is to obtain trade entry/exit points by identifying areas of the grid that the energy usage cost (the red line in subplot 6-d) is above the calculated optimized bid-offer (the dashed black line in subplot 6-d).} Residential load demand within this window is to be satisfied either using direct solar power and/or battery power. The DER is prepared to support this transaction by ensuring that enough energy is available to meet system constraints.
 If the battery SOC is greater than 60\%, this indicates that the battery has been under-utilized and the simulation would execute another iteration by reducing the offer cost which in turn achieves better peak load shavings and net earnings. Similarly, if the EV final SOC conditions are not satisfied, this would indicate that sum of window lengths must be shortened to enable more time for EV charging.



    
\section{Conclusion and Future Scope}\label{sec:conclusion}

{This paper proposed a decentralized Blockchain market design and optimization architecture that reduces peak load demand by integrating DER.} The Blockchain architecture is divided into two modules, namely the secure Blockchain-based P2P energy trading and intelligent energy prediction model to fulfill energy demand in a distributed network.
The Blockchain strategy determines optimal entry and exit energy trading window using user inputs, environmental information, and supervised learning {to maximize earnings and prepare DERs to support the transaction.} 
 {The Blockchain strategy helps providing DR and reducing overall energy consumption.} 

Future work can be conducted on new trading mechanisms. This includes expanding the study to include supervised and machine learning algorithms to better predict market trends, bi-directional vehicle to grid energy flow compatibility, Internet of things (IoT) integration to communicate energy demand/source levels for optimized energy storage, and finally to accommodate inter-community trading along with ancillary services to the grid.




\bibliographystyle{ieeetr}
\bibliography{sample}


\end{document}